# Science Advances

# Manuscript Template

FRONT MATTER

## Title

Efficient Learning of Mixed-State Tomography for Photonic Quantum Walk

## Authors

Qin-Qin Wang[1,2], Shaojun Dong[3], Xiao-Wei Li[4], Xiao-Ye Xu[1,2,5,*], Chao Wang[3], Shuai Han[6], Man-Hong Yung[7,†], Yong-Jian Han[1,2,3,5,‡], Chuan-Feng Li[1,2,5,§], and Guang-Can Guo[1,2,5]

## Affiliations

[1]CAS Key Laboratory of Quantum Information, University of Science and Technology of China, Hefei 230026, China.

[2]CAS Center for Excellence in Quantum Information and Quantum Physics, University of Science and Technology of China, Hefei 230026, China.

[3]Institute of Artificial Intelligence, Hefei Comprehensive National Science Center, Hefei 230031, China.

[4]Department of Physics, Southern University of Science and Technology, Shenzhen 518055, China.

[5]Hefei National Laboratory, University of Science and Technology of China, Hefei 230088, China.

[6]Yangtze Delta Region Industrial Innovation Center of Quantum and Information Technology, Suzhou 215100, China.

[7]Institute for Quantum Science and Engineering, Southern University of Science and Technology, Shenzhen 518055, China.

*xuxiaoye@ustc.edu.cn
†yung@sustech.edu.cn
‡smhan@ustc.edu.cn
§cfli@ustc.edu.cn

## Abstract

Noise-enhanced applications in open quantum walk (QW) have recently seen a surge due to their ability to improve performance. However, verifying the success of open QW is challenging, as mixed-state tomography is a resource-intensive process, and implementing all required measurements is almost impossible due to various physical constraints. To address this challenge, we present a neural-network-based method for reconstructing mixed states with a high fidelity ($\sim 97.5\%$), while costing only $50\%$ of the number of measurements typically required for open discrete-time QW in one dimension. Our method employs a neural density operator that models the system and environment, followed by a generalized natural-gradient-descent procedure that significantly speeds up the training process. Moreover, we introduce a compact interferometric measurement device, improving the scalability of our photonic QW setup that enables experimental learning of mixed states. Our results demonstrate that highly expressive neural networks can serve as powerful alternatives to traditional state tomography.

**Teaser**

An efficient mixed-state learning method on the quantum-walk system is demonstrated with an accelerated training process.

**MAIN TEXT**

**Introduction**

Quantum walk (QW) provides a basic framework for developing effective quantum algorithms and simulating complex phenomena (1-4). Through interaction with a certain amount of noise, the open (i.e., noisy) QW can produce a remarkable improvement in quantum transport over the noise-free QW (5, 6). And such a noise-enhanced feature aids in problem-solving efficiency in tasks such as graph isomorphism testing (7), a maze escape (8, 9), and ranking elements (10) in large networks. While as a prototypical dynamical process, the open QW can model the dissipative evolution of quantum neural networks (11, 12), and such simulation enables better performance to process various issues like pattern recognition (13). Adding controlled noise into the quantum evolution, QW can be dynamically initialized in any high-dimensional form (14-17) and generate the Haar random unitary operators (18, 19) required for quantum computation. Implementing the open QW and validating these noise-assisted computational and simulated performances demand a complete density-matrix characterization. However, identifying such a system with inherent high-dimensional structures is indeed a fundamental task to be solved in quantum information science (20).

For a discrete-time QW (21), the usual local tomographic technique was experimentally demonstrated that enables one to access the two correlated subsystems of coin (22-25) and position (26, 27), respectively. The tomography was subsequently extended to the walker's complete wave-function with a pure-state hypothesis (28), or a low-rank state when decoherence should be estimated (29). An alternative tomographic method was recently developed with machine-learning techniques (30-32), where a trained neural network can be used to recognize and classify QW states (33,34). The complicated mapping from the measured data to the probability amplitudes can also be learned supervisedly (35), and a "physical" wave-function of QW is obtained by projecting the un-physical neural-network outputs (36). Nevertheless, these methods mainly focus on the tomography of a closed QW, and none of them could be applicable for the mixed density-matrix reconstruction of an open QW, whose number of parameters scales quadratically with the pure-state one.

Here, we experimentally realize the full density-matrix tomography of a one-dimensional (1D) discrete-time QW with arbitrary mixing. We achieve this by parameterizing the complex-valued matrix elements using a neural density operator (NDO) (37), because such an effective ansatz has been shown to have better performance than standard maximum likelihood estimation in terms of the reconstruction fidelity and the number of measurements required. For training NDO to maximize the measurement data likelihood more efficiently, the generalized natural gradient descent (GNGD) procedure, recently developed in our group (38), is adopted to achieve about an order of magnitude of convergence rate than traditional gradient descent procedure. The effective mixed-state tomography of open QW, with a lower requirement of the number of measurements and training iterations for high-fidelity reconstructions, is benchmarked on synthetic datasets. Moreover, we further experimentally demonstrate that the trained NDO can learn noisy quantum states from partial measurements on a photonic QW. And the amount of mixing of the QW caused by the interaction with the environment can be well captured.

## Results

### Neural-network ansatz:

The quantum state of open systems is generally described by the density matrix $\rho = \sum_{v,v'} \rho(v, v') |v\rangle \langle v'|$ in a basis $|v\rangle$. To obtain the complex-valued matrix elements, a tomography with a series of measurements on a collection of bases $\{|v^n\rangle\}$ is necessary. The substantial demand for the number of complete tomographic measurements and a complicated data post-processing process make it impractical for a complex system (20), whereas neural-network-assisted tomography enables high-fidelity reconstruction of quantum states with fewer measurement resources (39). The core step is an effective variational parametrization of the measurement distributions on a quantum state in terms of a neural-network model to infer the complex-valued probability amplitudes (40), and the density matrix elements that rely on a linear reconstruction from the trained model distributions (41). The linear reconstruction of $\rho$ lacks the positivity constraint such that it cannot give a physical quantum state. To give a physical output, an NDO ansatz can be constructed by purifying the mixed state using an auxiliary Hilbert space $\mathcal{H}_a$ to ensure the composite quantum state in the enlarged space $|\Psi\rangle = \sum_{va} \Psi(v, a) |v\rangle |a\rangle$ is pure (37). Then, the mixed state of open systems can be written as a reduced state of the composite wavefunction, that is, $\rho = \text{Tr}_a[|\Psi\rangle\langle\Psi|]$. The traditional restricted Boltzmann machine as an ansatz of the wavefunction of the physical system is a two-layer neural network (39), including a visible layer $v$ connected to a hidden one $h$ (shown by the upper red region in Fig. 1(A)). Here, to express the composite wavefunction $|\Psi\rangle$ in an enlarged space, an additional ancillary layer $a$ is imposed as the auxiliary Hilbert space $\mathcal{H}_a$ for purifying the mixed state $\rho$, which corresponds to a fictitious environment mixed with the open systems (herein open QW), as shown by the bottom purple region in Fig. 1(A). Then, the probability amplitude of the composite wavefunction $\Psi(v, a)$ can be parameterized by the three-layer network:

$$\Psi_{\lambda\mu}(v, a) = Z_\lambda^{-\frac{1}{2}} \sqrt{p_\lambda(v, a)} e^{i \log p_\mu(v, a)/2}, \tag{1}$$

where $p_\theta(v,a) = e^{\sum_i \log(1+e^{W_\theta^{[i]}v+c_\theta^{[i]}}) + a^\mathrm{T} U_\theta v + b_\theta^\mathrm{T} v + d_\theta^\mathrm{T} a}$ ($\theta = \{\lambda, \mu\}$) is the joint distribution of the visible and ancillary layers, mimicking the amplitude and phase of the wavefunction. $A^{[i]}$ denotes $i$-th row of $A$, and $A^\mathrm{T}$ is the transpose of $A$. Vectors $b_\theta$, $c_\theta$, and $d_\theta$ are the biases coupled to the visible, hidden, and ancillary neurons, respectively. $Z_\lambda$ is a normalization constant. The NDO ansatz of the density matrix of the open system can be obtained by tracing out the auxiliary system (42-44):

$$\rho_{\lambda\mu} = \sum_{v,v'} \left[ \sum_a \Psi_{\lambda\mu}(v,a) \Psi^*_{\lambda\mu}(v,a) \right] |v\rangle\langle v'|. \tag{2}$$

Thus, the state tomography is mapped into an unsupervised learning task for training the network parameters such that the trained NDO gives an approximate physical state, i.e., $\rho_{\lambda\mu} \simeq \rho$. The expressive ability of the NDO ansatz depends on the number of the hidden and ancillary neurons. We now show how to accomplish the NDO tomography for an open QW.

**Open quantum walk model:**

Herein, we consider a 1D discrete-time QW that is composed of two interacting subsystems of the coin and the lattice, in which the probability amplitude of a quantum walker spreads between lattice sites on a line depending on the internal coin states (21). The open version of QW taking into account the effects of the ambient environment is shown in Fig. 1(B). The quantum state of open QW is described by a density matrix acting on the product Hilbert space $\mathcal{H} \equiv \mathcal{H}_C \otimes \mathcal{H}_l$, where $\mathcal{H}_C \equiv \mathrm{span}\{|s\rangle : s = \uparrow, \downarrow\}$ and $\mathcal{H}_l \equiv \mathrm{span}\{|l\rangle : l \in \mathbb{Z}\}$ represent the coin and the lattice subspace, respectively. Then the reference base is a tensor product of the basis for each subsystem $|v\rangle = |s\rangle|l\rangle$. And the density matrix on this base takes the form:

$$\rho = \sum_{s,l;s',l'} \rho_{sl,s'l'} |s\rangle|l\rangle\langle s'|\langle l'|. \tag{3}$$

The dynamics of open QW at each discrete step $t$ can be given by (45): $\rho_{t+1} = \sum_k \hat{E}_k \rho_t \hat{E}_k^\dagger$, in terms of the Kraus operators $\hat{E}_0 = \sqrt{1-(w_s+w_l)}\hat{U}$, $\hat{E}_1 = \sqrt{w_s}\hat{\mathbb{P}}_s\hat{U}$ and $\hat{E}_2 = \sqrt{w_l}\hat{\mathbb{P}}_l\hat{U}$. $\hat{E}_0$ contributes to the coherent evolution part and the unitary operator $\hat{U} = \hat{S}\hat{R}$, where conditional shift operator $\hat{S} = \sum_l(|\uparrow\rangle\langle\uparrow| \otimes |l+1\rangle\langle l| + |\downarrow\rangle\langle\downarrow| \otimes |l\rangle\langle l|)$ and coin-flip operator $\hat{R}(\alpha) = e^{-i\alpha\hat{\sigma}_y}\hat{\sigma}_z$. $\hat{\sigma}_y$ and $\hat{\sigma}_z$ are the Pauli matrices. The ensemble of $\hat{E}_1$ and $\hat{E}_2$ plays a role in eliminating coherence within the physical system, presented in the off-diagonal terms of the density matrix $\rho$. Projectors $\hat{\mathbb{P}}_s = \sum_l |s,l\rangle\langle s,l|$ and $\hat{\mathbb{P}}_l = \sum_s |s,l\rangle\langle s,l|$ project the walker into the coin state $|s\rangle$ and lattice state $|l\rangle$, respectively. The mixing parameter $w = w_s + w_l \in [0,1]$ quantifies the transition from the closed QW to the classical random walk (CRW). After an $N$-step walk, the size of the high-dimensional lattice state $|l\rangle$ is $N+1$. Considering the coin subspace is modeled as a two-level system, the total dimension of $\rho$ is thus $4(N+1)^2$. Here, we utilize the NDO ansatz $\rho_{\lambda\mu}$ in Eq. (2) as the variational representation of the target QW state $\rho$ in Eq. (3), which is to be reconstructed. Specifically, the visible layer encodes the base $|v\rangle$ of dimension $2(N+1)$ for an $N$-step open QW. The size of the ancillary layer determines the dimension of the fictitious environment Hilbert space used to purify the target QW state. The larger number of hidden and ancillary neurons are needed enabling the NDO ansatz to capture the increasingly complex correlations within the open QW and between the system and its environment, respectively, as the number of time steps $N$ and the mixing parameter $w$ increases. Given a series of measurement datasets $\{P(v^n)\}$ on multiple bases $\{|v^n\rangle\}$ imposed by the target QW state, the network parameters $\{W_{\lambda\mu}, U_{\lambda\mu}, b_{\lambda\mu}, c_{\lambda\mu}, d_{\lambda\mu}\}$ are trained on the datasets such that the reconstructed model distributions approximate the target measurement distributions, i.e., $\{P_{\lambda\mu}(v^n)\} = \{\rho^n_{\lambda\mu}(v^n, v^n)\} \simeq \{P(v^n)\}$. The density matrix $\rho^n_{\lambda\mu} = \hat{\mathcal{U}}(v^n, v)\rho_{\lambda\mu}\hat{\mathcal{U}}^\dagger(v^n, v)$ on the base $|v^n\rangle$ is constructed through the base transformation. With the optimized network parameters substituted into Eq. (2), the target mixed state of open QW can be reconstructed by the NDO, i.e., $\rho_{\lambda\mu} \simeq \rho$.

**Experimental implementation:**

QW can be realized on multiple physical platforms (46), and our experimental setup for photonic QW is shown in Fig. 2. More details are given in the Methods. In the experiment, we adopted the heralded single photons as the walker. The coin $\{|\uparrow\rangle, |\downarrow\rangle\}$ and the lattice $\{|l\rangle\}$ are encoded in the two polarizations $\{|H\rangle, |V\rangle\}$ and the different arrival times of photons $\{|t_l\rangle\}$, respectively. Polarization flip $\hat{R}(\alpha)$ is realized by tuning the angle of half-wave plates. A calcite crystal delays $|V\rangle$ by one time-bin length ($\Delta t \sim 5\,\mathrm{ps}$) relative to $|H\rangle$, which realizes the shift operator $\hat{S}$ on the time-bin modes. Considering each state $|t_l\rangle$ in the time basis is represented by a photonic wave packet with a width of $\sim 140$ fs that is far less than the time interval $\Delta t$, thus the overlap between different time bins is negligible. Through cascading $N$ times the two operators of $\hat{R}$ and $\hat{S}$, one can implement an $N$-step walk dynamics. The outgoing $N+1$ time bins are then injected into a Michelson interferometer for measurement base transformation, followed by an upconverted single-photon detector.

The Michelson interferometer in Fig. 2(A3) is a core ingredient for full state tomography and eases the usual demand of constructing a duplicate QW system (see lower panel in Fig. 2(B)) to obtain the phase relation between lattice sites (22). A moveable mirror in the interferometer introduces a controllable time delay between the two arms, which makes the photons with horizontal polarization $|H\rangle$ travel the integer multiples of 5 ps faster (or later) than those with vertical polarization $|V\rangle$. After the interferometer, a polarization analyzer with a QWP-HWP-PBS setting performs the single-qubit Pauli measurements on $\hat{\sigma}_x$-, $\hat{\sigma}_y$- and $\hat{\sigma}_z$-basis. Therefore, the interferometric measurements between all time bins here can be viewed as the system's projections on a series of basis $\{|\uparrow\rangle|l\rangle\}$, $\{|\downarrow\rangle|l\rangle\}$, $\{\frac{1}{\sqrt{2}}(|\uparrow\rangle|l\rangle \pm i|\downarrow\rangle|l'\rangle)\}$, and $\{\frac{1}{\sqrt{2}}(|\uparrow\rangle|l\rangle \pm |\downarrow\rangle|l'\rangle)\}$ ($l, l' = 0, 1, \cdots, N$). Moreover, even with the number of steps $N$ increasing, the required amounts of the optical elements for these interferometric measurements remain constant thanks to the compact structure. Consequently, the compact interferometer can save 50% of optical resources as there is no need to build another duplicate QW to implement these measurements.

One can classify these projection measurements into a set of bases $\{|v^n\rangle\}$. The first base $|v^0\rangle$ is the reference base $|v\rangle = \{|\uparrow\rangle|l\rangle, |\downarrow\rangle|l\rangle\}_{l=0}^N$, while the bases $|v^{2k-1}\rangle$ and $|v^{2k}\rangle$ ($k \in \{1, 2, \cdots, N+1\}$) are written as $\{\frac{1}{\sqrt{2}}(|\uparrow\rangle |l\rangle \pm i|\downarrow\rangle |[l-(k-1)]\mathrm{mod}(N+1)\rangle)\}_{l=0}^{N}$ and $\{\frac{1}{\sqrt{2}}(|\uparrow\rangle |l\rangle \pm |\downarrow\rangle |[l-(k-1)]\mathrm{mod}(N+1)\rangle)\}_{l=0}^{N}$, respectively. Note that the dividend in the mod function could be negative integers when $l < k-1$, so that the mod function can give values between 0 and $N$ but arrange them in different orders for different $k$. Base transformation matrix $\hat{\mathcal{U}}(v^n, v)$ consists of a conditioned shift operator with the periodic boundary ( $N+1$ -cycle) $\hat{S}' = \sum_{l=0}^{N} |\uparrow\rangle \langle\uparrow| \otimes |l\rangle \langle l| + |\downarrow\rangle \langle\downarrow| \otimes |(l-1)\mathrm{mod}(N+1)\rangle \langle l|$ and two elementary gates rotating the reference base $\hat{\sigma}_z$ in coin space to the local base $\hat{\sigma}_x$ and $\hat{\sigma}_y$, respectively. Then, the bases $|v^{2k-1}\rangle$ and $|v^{2k}\rangle$ can be obtained using the conditioned shift operator $(\hat{S}')^{k-1}$ together with two elementary gates to act on the reference base $|v\rangle$. We note that for an $N$-step QW, $N$ times the conditioned shift operator $\hat{S}'$ is needed to achieve interferometric measurements between all lattice sites of size $N+1$.

Throughout the experiment, the probability distribution $P(v^n)$ in each base $|v^n\rangle$ is analyzed by utilizing the up-converted detector in Fig. 2(A4). With a spatial delay line, one can scan a strong pump light of 300 mW to upconvert the signal photons in all time bins with an interval of 5 ps. The up-converted photons are filtered by a spectrum filter that consists of a dispersion prism in the 4-f system to reduce extra scattering noise. And the filtered photon counts are then measured by a PMT, with which we can obtain experimental distributions $P_{\mathrm{exp}}(v^n)$. Note that, using the linear reconstruction directly (41, 47), one can only access $2(N+1)^2 + 2(N+1)$ matrix elements according to the measurements on these $\mathcal{N}_b = 2(N+1) + 1$ sets of bases. Herein, we resort to the effective NDO method to learn the full $4(N+1)^2$ matrix elements from the incomplete measurements, which is an extension of learning some particular properties of quantum states using partial measurements (48, 49).

**Performance of the efficient tomography for QW:**

Before training the neural network with noisy experimental results, we start by benchmarking the NDO tomography of QW using synthetic datasets (50, 51). For each base $|v^n\rangle$, we define a family of $2(N+1)$ projectors $\{\hat{\mathbb{P}}_j^n = |v_j^n\rangle\langle v_j^n|\}_j$ that satisfy the normalization condition $\sum_j \hat{\mathbb{P}}_j^n = \mathbb{I}$ ($\mathbb{I}$ is the identity matrix). The projection distribution imposed on the target QW state $\rho$ can be obtained by the linear expression $P(v^n) = \{\mathrm{Tr}[\hat{\mathbb{P}}_j^n \rho]\}_j$, with $\sum_{v^n} P(v^n) = 1$. For an $N$-step QW, we generate the synthetic datasets $\{P(v^n)\}$ in a collection of $\mathcal{N}_b = 2(N+1) + 1$ bases $\{|v^n\rangle\}$ to train the neural network. The training process minimizes the cost function, defined as the total statistical distance $\mathcal{D}_\theta = \sum_n \sum_{v^n} P(v^n)\log[P(v^n)/P_\theta(v^n)]$, between the target and the reconstructed distributions. The process starts with the network parameters being initialized to random values, and in each optimization iteration $i$, the parameters are updated according to the GNGD procedure (38): $\theta_{i+1} = \theta_i - \eta_i G^{-1} \nabla \mathcal{D}_\theta$, where the searching step $\eta_i$ is determined by the line search process. To determine a well-performed metric $G$ that can speed up the gradient-based optimization, the core idea of GNGD is to introduce a proper reference space, which is regarded as a flat space here. By choosing the identity matrix as the metric of reference space $G_{ij}^{\mathrm{ref}} = \delta_{i,j}$, the metric for the cost function $G$ can be obtained through the conversion of coordinates (See Methods for details). All the NDO reconstructions given below were obtained in this way.

All the initial state is set to be $\rho_0 = |\psi_0\rangle\langle\psi_0|$ with $|\psi_0\rangle = \frac{1}{\sqrt{2}}(|\uparrow\rangle + i|\downarrow\rangle) \otimes |0\rangle$. We first focus on the simplest model where the flip angle $\alpha$ is time-independent and the mixing parameter $w = 0$. Consider that the coin operator is a Hadamard gate ($\alpha = \pi/4$), and the corresponding walk is the typical Hadamard QW. The fidelity of the NDO tomography performed on the synthetic datasets is reported by the green solid line in Fig. 3(A), and a uniformly high fidelity is achieved even with an increasing number of time steps. In a bit complicated case, the angle $\alpha \in [0, \pi)$ is set to be time-dependent and becomes completely random for each step. For a given step of this coherent disordered QW (52, 53), 100 disordered configurations are generated, corresponding to 100 different QW dynamics of the quantum states. The red dashed line in Fig. 3(A) shows the mean performance of the reconstruction fidelity obtained by averaging over the results of the 100 samples for each time step. A reconstruction fidelity error below $4 \times 10^{-3}$ was found for the coherent disordered walk with 3000 samples in total.

We then turn to a more challenging scenario of open QW interacting with a simulated environment (19, 54). Decoherence can be introduced by inserting an extra phase gate $\hat{R}(\beta) = e^{\frac{i}{2}\beta\hat{\sigma}_z}$ with a fast fluctuating phase $\beta \in [-\delta\beta, \delta\beta]$ into each step of coherent QW. The fluctuating degree $\delta\beta \in [0, \pi]$ can effectively add different degrees of decoherence like the mixing parameter $w \in [0, 1]$ in the open quantum-walk model because the results obtained for a certain value of the mixing parameter $w$ can be faithfully reproduced with a proper fluctuating degree $\delta\beta$ (50). Specially, $\delta\beta = 0$ corresponds to a coherent QW with $w = 0$, and $\delta\beta = \pi$ results in a fully decoherent CRW with $w = 1$. We also generated 100 samples of the open QW with different mixing for each time step, and the NDO reconstruction fidelity $F(\rho, \rho_{\lambda\mu}) = \text{Tr}(\sqrt{\sqrt{\rho}\rho_{\lambda\mu}\sqrt{\rho}})^2$ shown in Fig. 3(B) can still reach a high value of $\sim 97.5\%$ with the partial measurements. While this plot and our experiment below are for the example of the phase decoherence of the coin, we note that the NDO tomography is also valid for the open QW with depolarizing noise; see Sect. C of the Supplementary Materials (50). Beyond fidelity, an important question is whether the mixing of QW is reproduced accurately in the NDO reconstruction. The insets in Fig. 3(A, B) display the reconstructed error in the purity $|\text{Tr}[\rho_{\lambda\mu}^2] - \text{Tr}[\rho^2]|$ for the three scenarios. The mean value of the purity error is lower than $5 \times 10^{-3}$, $1 \times 10^{-2}$, and $5 \times 10^{-3}$ for the Hadamard, coherent disordered and open QWs, respectively, finding a close agreement with the expected results of the target state $\rho$.

All the training above for the open QW is implemented in a FORTRAN program, and the corresponding code can be found in Ref. (51). For the neural-network hyperparameters, the number of visible neurons is uniquely determined by the dimension of the basis $|v^0\rangle = |s\rangle|l\rangle$. The number of neurons in the hidden and visible layers is fixed to be 10 for each time step of the Hadamard QW and coherent disordered QW in Fig. 3(A) and 15 for the model of open QW in Fig. 3(B), respectively. In each iteration process, we directly sum over all the terms in $\nabla \mathcal{D}_\theta$ for being free of the sampling error of Monte Carlo sampling. The criterion used to stop training is that the norm of the gradient $\nabla \mathcal{D}_\theta$ reaches a fixed value of $10^{-8}$, and in general, the training stopped after 100-2000 iterations, depending on the dimension of the target density matrix and the choice of model hyperparameters of the neural network.

As a comparison, we performed standard maximum likelihood estimation (MaxLik) for the three types of QW. The MaxLik estimator we use takes the form: $\rho_\mathrm{m} = TT^\dagger/\mathrm{Tr}(TT^\dagger)$. The matrix $T$ is a lower triangular matrix with diagonal elements being real-valued and non-diagonal elements being complex-valued. For an $N$-step open QW, $4(N+1)^2$ real numbers need to be optimized such that the MaxLik can give a distribution $P_\mathrm{m}(v^n) = \{\mathrm{Tr}[\hat{\mathbb{P}}_j^n \rho_\mathrm{m}]\}_j$ approximately equal to the measurement one $P(v^n) = \{\mathrm{Tr}[\hat{\mathbb{P}}_j^n \rho]\}_j$ ($n = 0, 1, \cdots, 2(N+1)$) in terms of the total statistical distance as the cost function. The values of the $4(N+1)^2$ real numbers are randomly initialized and updated using the conjugate gradient (CG) algorithm (55). The stopping criterion is that the norm of the gradient reaches a fixed value of $10^{-8}$. The reconstruction fidelity and purity error for three types of QWs are shown in Fig. 3(C, D). We find that, for the three scenarios, the performance of the NDO learning method is clearly superior to that of the standard MaxLik method when the given measurement is partial, which is consistent with previous works (37, 62-64).

**Experimental mixed-state reconstructions:**

After benchmarking it on synthetic data, we now demonstrate the efficient learning of open QW with noisy experimental results. Using Polarizer1 in Fig. 2(A2), the walker's initial state is fixed to be a product state $\frac{1}{\sqrt{2}}(|H\rangle + i|V\rangle) \otimes |t_{l=0}\rangle$. The first experiment we performed was a five-step Hadamard walk interacting with a controllable simulated environment. Besides a Hadamard gate realized by an HWP with its optical axis oriented at $22.5°$, the extra phase gate $\hat{R}(\beta) = |H\rangle\langle H| + e^{i\beta}|V\rangle\langle V|$ is introduced by a configuration of QWP-HWP-QWP for each time step. The configuration features two QWPs rotated to an angle of $45°$ and a sandwiched HWP whose rotation angle controls the relative phase $\beta$ between horizontal and vertical polarization (56). To mimic the decoherence effect, the five sandwiched HWPs are installed on the motorized stages to constantly change their rotation angles with a controlled fluctuating degree $\delta\beta = 0, \pi/8, \pi/4, \pi/2, 3\pi/4$ and $\pi$. Then an ensemble measurement of $P_\mathrm{exp}(v^n)$ is performed on $\mathcal{N}_b = 13$ sets of bases by unitizing the Michelson interferometer (see Fig. 2(A3)) and the up-converted detector (see Fig. 2(A4)). As reported in Fig. 4(A), NDO can learn to reproduce the purity of open QW's mixed states present in the experimental data well. These results allow us to investigate the QW-to-CRW transition in greater depth than by only considering their transport behaviors (57) exhibited in the diagonal terms of the density matrix. For the Hadamard walk ($w=0$) shown in Fig. 4(B), beyond the typical characterization with pronounced side peaks, the underlying coherence is clearly presented in the complex-valued off-diagonal matrix elements of $\rho_{\lambda\mu}^\mathrm{exp}$. The decoherence introduced by the simulated environment can completely destroy the phase relation between lattice sites such that CRW ($w=1$) dominates. Consequently, it can bring the off-diagonal matrix elements to zero and cause the diagonal elements to have a classical binomial distribution, as shown in Fig. 4(C).

As shown in the inset of Fig. 4(a), the fidelity of the reconstructed full density matrix is greater than 0.919, and is superior to the one using the MaxLik method (>0.835). The deviation in the reconstruction fidelity between the experimental and synthetic datasets mainly results from systematic errors present in the experiment. The two experimental errors consist of an inexact orientation of the wave plates with a precision of $\pm 0.2°$ with respect to the correct angle and an inexact calibration of the QW interference network step-by-step with an accumulated phase error of $\pm 0.01\pi$ per step. We note that when the fluctuating degree $\delta\beta = 0$ and $\pi$, the target QW state $\rho$ is a rank-one density matrix and a completely mixed state, respectively. NDO only needs to learn the correlation within the physical system (captured by the hidden layer) or the correlation between the system and the ambient environment (captured by the ancillary layer). By contrast, when $\delta\beta = \pi/2$, NDO needs to learn both the complex correlations within the physical system and between the system and the ambient environment. This could lead to the reconstruction fidelity at $\delta\beta = \pi/2$ being slightly lower than the scenarios with $\delta\beta = 0$ and $\pi$.

As the total QW steps $N$ increase, the learning of mixed states with the dimension of $4(N+1)^2$ demands the complicated training of a larger neural network to keep its high expressive power. Thus, we introduce the GNGD procedure (38) to enhance the training efficiency of NDO. To better demonstrate the performance of GNGD for addressing complex networks of NDO, we extend the number of discrete time steps $N = 30$, and a thirty-step Hadamard walk is experimentally realized. Interferometric measurements on $\mathcal{N}_b = 63$ sets of bases are performed, and the obtained distributions $P_{\exp}(v^n)$ are displayed in Fig. 5(A). The mean value of classical indicator similarity of the measured distributions, defined as $\mathcal{S} = \sum_n \sum_{v^n} \sqrt{P_{\exp}(v^n)P_{\text{th}}(v^n)}/\mathcal{N}_b$, reads $0.965 \pm 0.008$, giving good agreement with the theoretical predictions. In Fig. 5(B), we compare the cost function $\mathcal{D}_\theta$ as a function of the number of iterations using gradient descent (GD), conjugate gradient (CG) (55), L-BFGS (58), and GNGD methods for training NDO, respectively. It can be found that the GNGD optimizer enables a lower value of $\mathcal{D}_\theta$ in one order of magnitude fewer iterations than the traditional GD-based optimization, and achieves a 5-fold speedup in the training process over CG and L-BFGS. As shown in Fig. 5(C), the GNGD-enhanced NDO can efficiently output the experimental mixed state $\rho_{\lambda\mu}^{exp}$ with $62^2$ complex-valued matrix elements, using fewer training iterations and partial measurements. The emblematic side peaks and the off-diagonal elements that represent coherence are still clear after the 30-step Hadamard walk. The reconstructed fidelity and purity are $0.789 \pm 0.011$ and $0.675 \pm 0.013$, respectively. A test on synthetic data yields the values of fidelity and purity both greater than $0.995$, indicating that the uncontrolled decoherence introduced by the real environment is the major deviation for the experimental reconstruction. While classical similarity indicates how much two distributions overlap, the fidelity we obtained is a true estimator of the quality of the quantum states produced, characterizing the quality of experimental quantum systems more accurately.

**Discussion**

In this work, we have experimentally demonstrated an effective learning of mixed states for open photonic QW in terms of the number of training iterations, the amount of measurement resources, and the reconstruction fidelity. The learning method is achieved by training a neural-network-parameterized density operator on measurement datasets and then outputting complex-valued matrix elements accelerated by the GNGD procedure. Although several works have already demonstrated the power of various neural networks for learning the mixed state of open systems, these proposals are mainly focused on the reconstruction of a single qubit (59), a single qudit (60), and the dominant eigenstates of a low-rank mixed state (61). Especially, the effectiveness of NDO tomography was demonstrated only in a simple two-qubit open system (37). Until recently, the tomography of the density matrix of dimensions up to $32^2$ has been reported using the restricted Boltzmann machine (62) and the conditional generative adversarial network (63, 64). By contrast, we both theoretically and experimentally realized the NDO tomography of a $2 \times n$ open QW system and the reconstruction of the mixed state with the largest number of matrix elements, up to $62^2$, surpassing all previously reported results. Furthermore, this experimental study is also the first time to verify the effectiveness of our self-developed GNGD optimization algorithm (38) for accelerating training on the long-standing challenging task of experimental learning of open QW. The GNGD optimizer can be applied to various neural-network architectures for effectively speeding up the gradient-based tomography of the quantum state (65, 66) and the quantum process (67-69).

In addition to the neural-network ansatz, an effective mixed-state tomography has recently been reported using the tensor-network ansatz (70). Compared to the locally purified state ansatz there, the correlations captured by the hidden and ancillary layers in NDO are inherently nonlocal. Therefore, the NDO ansatz we use is better suited to describe the $2 \times n$-dimensional open QW with inherent high-dimensional structures (39). To capture the complex correlations within the physical system and the environment more accurately as the time step increases, we usually increase the number of hidden and ancillary neurons, respectively. One can introduce the deep restricted Boltzmann machine with improved representational power (71), at the cost of growing training complexity, to further develop the NDO method to learn the open quantum systems. Adding an extra noise layer (40) or using global measurements (72, 73) could improve the robustness of state reconstructions against experimental errors. Overall, our approach provides a promising avenue for addressing the challenges associated with verifying open QW. We expect that these results will lead to new insights and discoveries in this exciting area of research on numerous physical systems such as fiber loop (74), spatial path (75), orbital angular momentum (76), transverse momentum (77, 78), hybrid architecture (79), etc. Moreover, the full quantum state tomography technique can be combined with the known abilities of arbitrary initialization (14, 24) and flexible manipulation, which can inspire a prospective QW platform for developing novel applications in a range of fields.

**Materials and Methods**

### Training the neural network by GNGD:

The training of the neural network is to learn the optimal parameters $\theta = \{\lambda, \mu\}$ that minimize the cost function $D_\theta$. The derivative of the cost function with respect to the network parameters reads:

$$\nabla D_\theta = -\sum_n \sum_{v^n} \frac{P(v^n)}{\rho_\theta^n(v^n,v^n)} \sum_{\alpha,\beta} \mathcal{U}^n(v^n,\alpha) \rho_\theta(\alpha,\beta) \nabla A_\theta(\alpha,\beta) \mathcal{U}^{n,\dagger}(\beta,v^n) + \mathcal{N}_b \sum_v \rho_\theta(v,v) \nabla A_\theta(v,v), \quad (4)$$

where $\rho_\theta^n(v,v') = \sum_{\alpha,\beta} \mathcal{U}^n(v,\alpha) \rho_\theta(\alpha,\beta) \mathcal{U}^{n,\dagger}(\beta,v')$, $\mathcal{U}^n = \langle v^n | v \rangle$ is the base transformation matrix from the reference base $|v\rangle$ to $|v^n\rangle$, $\mathcal{N}_b$ is the total number of

measurement bases. $\rho_{\lambda\mu}(\alpha,\beta) = Z_\lambda^{-1} e^{A_{\lambda\mu}}$ is the rebuild density matrix of the system, where

$$\begin{aligned}
A_{\lambda\mu}(v,v') &= \Gamma_\lambda^+(v,v') + i\Gamma_\mu^-(v,v') + \Pi_{\lambda\mu}(v,v'), \\
\Gamma_\theta^\pm(v,v') &= \frac{1}{2}[\sum_i \log(1 + e^{W_\theta^{[i]}v + c_\theta^{[i]}}) \pm \sum_i \log(1 + e^{W_\theta^{[i]}v' + c_\theta^{[i]}}) + b_\theta^T(v \pm v')], \\
\Pi_{\lambda\mu}(v,v') &= \sum_i \log\{1 + \exp[\frac{1}{2}U_\lambda^{[i]}(v+v') + \frac{i}{2}U_\mu^{[i]}(v-v') + d_\lambda^{[i]}]\}.
\end{aligned} \qquad (5)$$

The superscript $[i]$ means the $i$-th row elements. The network parameters to be optimal here are $\{W_\theta, U_\theta, b_\theta, c_\theta, d_\theta\}$. $W_\theta$ and $U_\theta$ are the weight matrices, whose dimension is $m \times d$ with $d$ being the degree of freedom of the physical system and $m$ being the number of hidden or ancillary neurons. $b_\theta$, $c_\theta$ and $d_\theta$ are the vectors of dimension $d$ representing the biases coupled to the visible, hidden, and ancillary neurons, respectively. To accelerate the convergence efficiency of training, the GNGD method emphasizes introducing a proper reference space (38). Here, the space span by the density matrix $\rho_\theta(\alpha,\beta)$ is selected as the reference manifold. We choose a simple identity matrix $G^{\text{ref}}_{\alpha,\beta;\alpha',\beta'} = \delta_{\alpha,\alpha'}\delta_{\beta,\beta'}$ as the metric of reference space. Through the conversion of coordinates, the metric $G$ for the cost function can be determined as:

$$G_{i,j} = \sum_{\alpha,\beta;\alpha',\beta'} \frac{\partial \rho_\theta(\alpha,\beta)}{\partial \theta_i} G^{\text{ref}}_{\alpha,\beta;\alpha',\beta'} \frac{\partial \rho_\theta(\alpha',\beta')}{\partial \theta_j} \qquad (6)$$

With this metric, the network parameters are updated according to the gradient-based optimization as:

$$\theta_{i+1} = \theta_i - \eta_i G^{-1} \nabla \mathcal{D}_\theta \qquad (7)$$

where $\eta_i$, which is determined by the line search process, is the searching step at iteration $i$.

**Heralded single-photon source:**

A Ti: sapphire laser source (Mira 900, Coherent) launches a series of optical pulses with a temporal pulse width of 140 fs, a repetition rate of 76 MHz, the wavelength $\lambda$=800 nm, and an average power of 1 W. The ultrafast optical pulses focused by Lens1 (focal length f=100 mm) pump the first $\beta - \text{BaB}_2\text{O}_4$ (BBO1), causing the second harmonic generation. The frequency-doubled pulses with horizontal polarization and the residual fundamental pulses with vertical polarization are collimated by Lens2 and then spatially separated by the first dichroic mirror (DM1). The fundamental pulses go through a pair of cylinder lenses (not shown) to reform the beam's profile into a Gaussian shape and are then reused for position-resolved state detection. The frequency-doubled pulses with $\lambda$=400 nm and an average power of 200 mW are used as a pump source and focused on the BBO2 by lens3 (focal length f=100 mm). Herein, BBO2 is designed for type-II, non-degenerate, and ``beam-like'' spontaneous parametric down-conversion (SPDC). The generated time-correlated photon pairs (namely, signal photons with $\lambda$ =780 nm and horizontal polarization, and idler photons with $\lambda$=821 nm and vertical polarization) are spatially separated by the non-collinear SPDC process with a half opening angle of $3°$. The idler photons, collimated by the Lens5 and cleaned by a spectrum filter, are coupled to a single-mode fiber and then detected by the avalanche photodiode detector (SPCM-AQRH-14-FC, Excelitas) to herald the appearance of the signal photons. The coincidence count rate between idler and signal photons is $4.5 \times 10^2$ pairs/(s mW). The collimated and filtered signal photons are adopted as the walker and sent to the quantum walk (QW) module.

**Time-multiplexing photonic quantum walk:**

The time-multiplexing QW is achieved by encoding the walker's lattice and coin space in the arrival time and polarization of signal photons, respectively. The signal photons at the input of the QW module are set to be in the first time bin, i.e., the origin in lattice space with $l = 0$. Then, the initial polarization state of signal photons $|\uparrow_y\rangle = \frac{1}{\sqrt{2}}(|H\rangle + i|V\rangle)$ is prepared through a polarization-dependent beam splitter (PBS), a half-wave plate (HWP), followed by a quarter-wave plate (QWP). $|H\rangle$ and $|V\rangle$ represent horizontal and vertical polarization, respectively. Thus, the initial state of the walker is $|\psi_0\rangle = |l = 0\rangle \otimes |\uparrow_y\rangle$. In the protocol of Hadamard QW, the unitary operator of a single step $\hat{U} = \hat{S}\hat{R}$ is composed of an HWP rotated to a specific angle at $22.5°$ (for operator $\hat{R}$) and a calcite crystal (for operator $\hat{S}$). Each piece of calcite is cut parallel to its optical axes to a length of 8.98 mm, whose birefringence $\Delta = 0.167$ at $\lambda$=800 nm will induce a 5ps time shift between the horizontal $|H\rangle$ and vertical polarization $|V\rangle$ of the signal photons. In the experiment, $N$ sets of HWPs and calcite crystals are positioned to realize an $N$-step Hadamard walk. Thus, the lattice space of the walker at the end of the QW module consists of the superposition of $N + 1$ time bins with a time interval of 5 ps. And for each lattice site, the walker has an internal coin state, which is usually distinct at a different site. A QWP-HWP-QWP setting for realizing the phase gate operator $\hat{R}(\beta) = e^{\frac{i}{2}\beta\hat{\sigma}_z}$ can be introduced at each QW step to mimic a controlled decoherence effect. Two QWPs are rotated to an angle at $45°$. HWP in the setting is installed on the motorized stage (PR50PP, Newport) to change its rotation angle constantly. The fluctuating range $[-\delta\beta, \delta\beta]$ determines the degree of decoherence and has a similar effect as the mixing parameter $w$: $\delta\beta = 0$ is the coherent QW with $w = 0$, and $\delta\beta = \pi$ results in a fully decoherent classical walk with $w = 1$.

**Base transformation and phase-locking technique:**

The base transformation is achieved by a Michelson interferometer consisting of a PBS, two QWPs rotated to an angle at $45°$, and two mirrors. One mirror is piezoelectric ceramic (PZT)-driven to compensate for the phase fluctuation of the interferometer, and the other mirror is installed on a motorized positioning system (KXL06100, Suruga). The moveable mirror is controlled to induce a time shift between the two arms of the interferometer, which is accurately the integer multiples of 5 ps for ensuring maximal visibility between different time bins of the signal photons and avoiding temporal mode mismatching. The phase difference between the two arms can be controlled by the phase shifter (PS) and locked in a fixed value by an ancillary Helium-Neon (He-Ne) laser using a proportional-integral-differential (PID) feedback unit (80). The PS is the configuration of QWP-HWP-QWP, where two QWPs are rotated to an angle at $45°$ and the rotation angle of HWP controls the relative phase between $|H\rangle$ and $|V\rangle$ states. Initially, the He-Ne laser source (HNL050LB, Thorlabs) with $\lambda$=632.8 nm is prepared in the state $|\downarrow_x\rangle = \frac{1}{\sqrt{2}}(|H\rangle - |V\rangle)$ through the Polarizer2 and then launched to the same propagation path with the signal

photons through DM2. The signal photons and the He-Ne laser pass through the interferometer and are spatially separated by DM3. The reflected He-Ne laser is projected in the $|\uparrow_x\rangle = \frac{1}{\sqrt{2}}(|H\rangle + |V\rangle)$ state and detected by a amplified Si photodetector (PDA8A/M, Thorlabs). The detected He-Ne laser intensity is initially set near a reference value through PS. Within the acquisition time of time-correlated photon pairs, the PID unit compares the detected intensity using the amplified Si photodetector to the reference value and then uses the difference to calculate a new input value through the PID algorithm that is designed to keep the detected intensity at the reference value. Thus, the unstable phase difference of the interferometer for signal photons can then be locked with the He-Ne laser. PS can be used to compensate the phase difference of the signal photons introduced by the two arms of the interferometer to zero. After passing through the Michelson interferometer and the polarization analyzer, the signal photons are collected into a 0.1m-long single-mode fiber (SMF) for spatial filtering. The two HWPs at the input and output of the SMF are used for polarization maintenance.

**Frequency up-conversion single-photon detector:**

To measure the probability distribution $P(v^n)$ in each base $|v^n\rangle$, a position-resolved detection of the signal photons is constructed. The lattice space of the signal photons consists of a pulse train with a time interval of 5 ps. Herein, we utilize a single-photon frequency upconversion system converting the high time resolution to the spatial resolution. Specifically, the residual fundamental pulses with λ=800 nm and average power 300 mW during the second harmonic generation serve as the strong pump pulse. The up-converted photons are generated in BBO3, where the signal photons and the strong pump pulse meet, and then detected by a photomultiplier tube (H10682, Hamamatsu). In the up-conversion process, the propagation path of the strong pump pulse is delayed and aligned using two moveable mirrors installed on a motorized positioning system (KXL06100, Suruga). And the alignment of the signal photons is achieved by replacing them with a reference coherent laser to optimize the spatial mode matching. In addition, the delay of the strong pump pulse is fine-tuned to match each pulse in the signal-photon pulse train in the time domain. A spectrum filter composed of a dispersion prism is used to reduce the background.

48. J. Gao, L.-F. Qiao, Z.-Q. Jiao, Y.-C. Ma, C.-Q. Hu, R.-J. Ren, A.-L. Yang, H. Tang, M.-H. Yung, and X.-M. Jin, Experimental machine learning of quantum states, *Phys. Rev. Lett.* **120**, 240501 (2018).

49. M. Yang, C.-l. Ren, Y.-c. Ma, Y. Xiao, X.-J. Ye, L.-L. Song, J.-S. Xu, M.-H. Yung, C.-F. Li, and G.-C. Guo, Experimental simultaneous learning of multiple nonclassical correlations, *Phys. Rev. Lett.* **123**, 190401 (2019).

50. See Supplemental Materials for a more discussion on the neural-network ansatz, neural-network reconstruction for the open quantum walk, benchmarking neural-network reconstruction of an open QW with depolarizing noise, and relation between the fluctuating degree and the mixing parameter.

51. Codes for generating the synthetic datasets and the quantum state tomography using the neural density operator are available at https://github.com/Shaojun-Dong/NDO.

52. A. Geraldi, A. Laneve, L. D. Bonavena, L. Sansoni, J. Ferraz, A. Fratalocchi, F. Sciarrino, A. Cuevas, and P. Mataloni, Experimental investigation of superdiffusion via coherent disordered quantum walks, *Phys. Rev. Lett.* **123**, 140501 (2019).

53. Q.-Q. Wang, X.-Y. Xu, W.-W. Pan, K. Sun, J.-S. Xu, G. Chen, Y.-J. Han, C.-F. Li, and G.-C. Guo, Dynamic-disorder-induced enhancement of entanglement in photonic quantum walks, *Optica* **5**, 1136-1140 (2018).

54. A. Schreiber, K. N. Cassemiro, V. Potoček, A. Gábris, I. Jex, and C. Silberhorn, Decoherence and disorder in quantum walks: From ballistic spread to localization, *Phys. Rev. Lett.* **106**, 180403 (2011).

55. Y. S. Tao, Numerical Estimation Schemes for Quantum Tomography, *arXiv*:1302.3399 [quant-ph] (2013).

56. B.-G. Englert, C. Kurtsiefer, and H. Weinfurter, Universal unitary gate for single-photon two-qubit states, *Phys. Rev. A* **63**, 032303 (2001).

57. T. A. Brun, H. A. Carteret, and A. Ambainis, Quantum to classical transition for random walks, *Phys. Rev. Lett.* **91**, 130602 (2003).

58. D. C. Liu and J. Nocedal, On the limited memory bfgs method for large scale optimization, *Mathematical Programming* **45**, 503-528 (1989).

59. Y. Zuo, C. Cao, N. Cao, X. Lai, B. Zeng, and S. Du, Optical neural network quantum state tomography, *Adv. Photonics* **4**, 026004 (2022).

60. D. Koutný, L. Motka, Z. c. v. Hradil, J. Řeháček, and L. L. Sánchez-Soto, Neural-network quantum state tomography, *Phys. Rev. A* **106**, 012409 (2022).

61. A. Melkani, C. Gneiting, and F. Nori, Eigenstate extraction with neural-network tomography, *Phys. Rev. A* **102**, 022412 (2020).

62. E. S. Tiunov, V. V. T. (Vyborova), A. E. Ulanov, A. I. Lvovsky, and A. K. Fedorov, Experimental quantum homodyne tomography via machine learning, *Optica* **7**, 448–454 (2020).
*Science Advances*     Manuscript Template     Page **17** of **24**

**Acknowledgments:**

We thank Zhih-Ahn Jia, Hai-Lan Ma and Zi-Wei Cui for useful discussions.

**Funding:**

This work was supported by Innovation Program for Quantum Science and Technology (No. 2021ZD0301200), National Natural Science Foundation of China (Nos. 12022401, 62075207, 11874343, 12104433, 11821404), the Fundamental Research Funds for the Central Universities (No. WK2470000030), and the CAS Youth Innovation Promotion Association (No. 2020447). Q.-Q. Wang acknowledges support from China Postdoctoral Science Foundation (No. 2021M703108), Fundamental Research Funds for the Central Universities (No. WK2030000081) and National Natural Science Foundation of China (No. 12204468).

**Author contributions:**

Q.-Q. W. and S. D. contributed equally to this work. Q.-Q. W. performed the experiments under the guidance of X.-Y. X. and C.-F. L. S. D. performed the theoretical works under the guidance of Y.-J. H. and C. Wang. X.-W. L., S. H. and M.-H. Y. provided theoretical supports. Q.-Q. W. and S. D. drafted the manuscript, with revisions from M.-H. Y., X.-Y. X., Y.-J. H. and C.-F. Li. C.-F. L. and G.-C. G. supervised the project.

**Competing interests:**

Authors declare that they have no competing interests.

**Data and materials availability:**

All data is available in the main text and the Supplementary Materials.

**Figure captions**

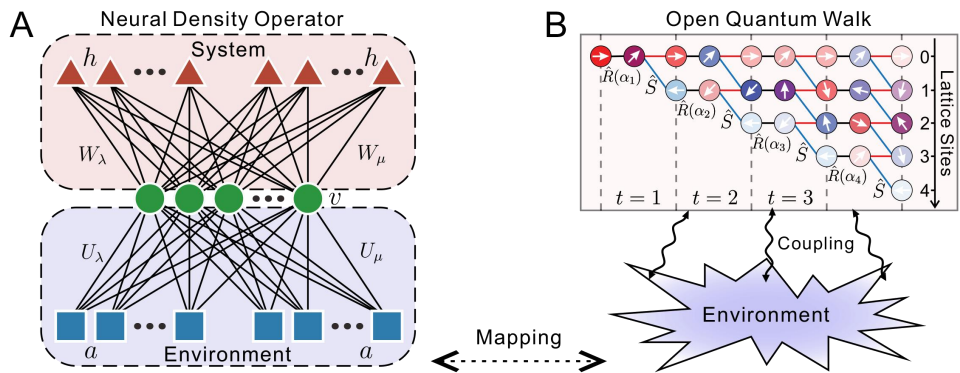

**Fig. 1. A schematic diagram of NDO and open QW. (A)** $|v\rangle$ is encoded in the visible layer (green circles), a hidden layer $h$ (red triangles) captures the correlation within the physical system, and an ancillary layer $a$ (blue squares) encodes the coupling between the system and the environment. Matrices $W_{\lambda,\mu}$ and $U_{\lambda,\mu}$ are the connecting weights. **(B)** Open QW consists of two degrees of freedom: the coin (white arrows) and lattice (colored circles), and the coupling environment introduces the mixing of this system.

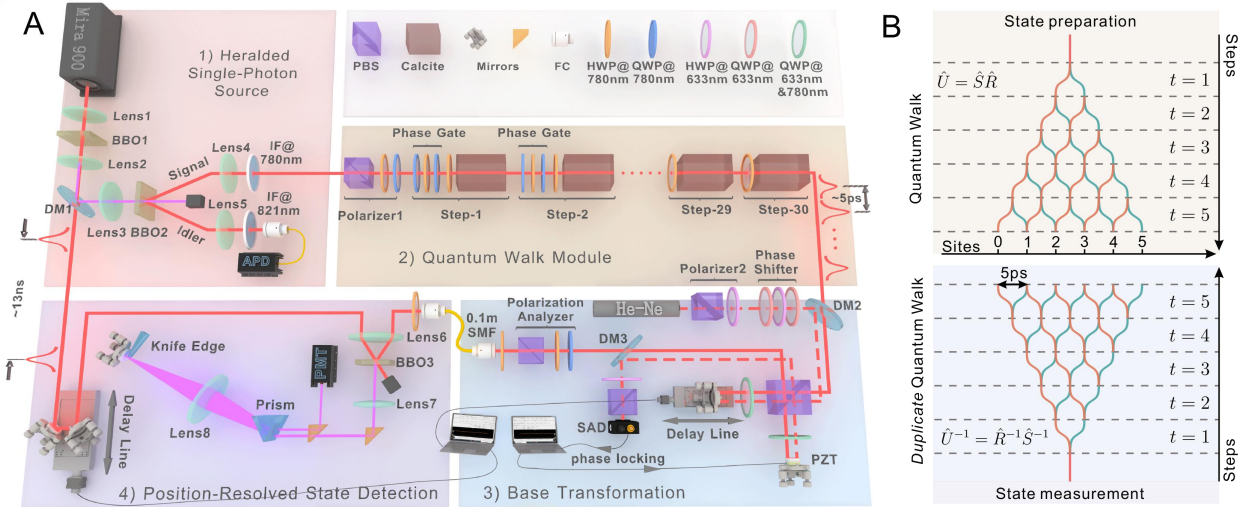

**Fig. 2. Photonic open QW. (A)** The experimental setup mainly has four central parts: 1) Spontaneous parameter down-conversion generates the time-correlated photon pairs, where signal photons as the walker and the idler photons serve to herald; 2) The open QW, as reported in the upper panel of **(B)**, involves the cascade of HWPs (for operator $\hat{R}(\alpha)$) and calcites (for operator $\hat{S}$) that realizes the unitary dynamics $\hat{U}^t$. An additional varying phase gate $\hat{R}(\beta)$ with the QWP-HWP-QWP setting is added to each time step to introduce incoherent contributions; 3) A Michelson interferometer together with a polarization analyzer carries out the base transformation $\hat{\mathcal{U}}(v^n, v)$ on the reference base $|v\rangle$, which mimics the effect of a duplicate QW in the lower panel of **(B)**. The PZT-driven mirror is applied for phase-locking by using a reference He-Ne laser, and another moveable mirror is used for adjustable arm length difference; 4) A single-photon frequency up-conversion implements the position-resolved detection of the walker in each base. **(B)** The schematic diagrams show the QW dynamics of a localized initial state (top) and the duplicate QW with time inversion constructed for the state measurements in different bases (bottom), respectively. A list of abbreviations: $\beta-\mathrm{BaB_2O_4}$ (BBO); dichroic mirror (DM); interference filter (IF); polarization-dependent beam splitter (PBS); half-wave plate (HWP); quarter-wave plate (QWP); piezoelectric ceramic (PZT); fiber collimator (FC); single-mode fiber (SMF); Si amplified detector (SAD); photomultiplier tube (PMT); avalanche photodiode detector (APD).

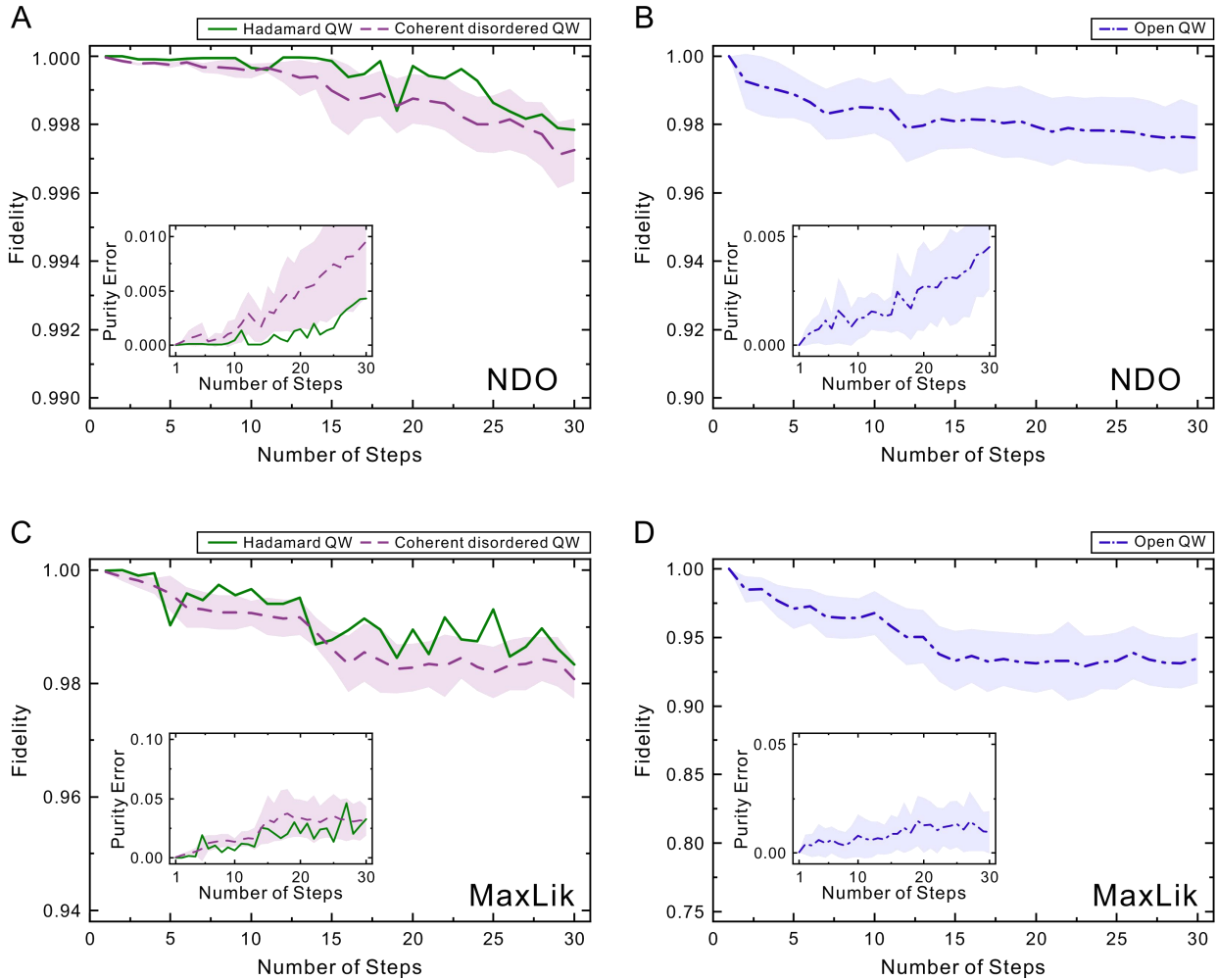

**Fig. 3. Benchmarking NDO and MaxLik tomography using partial measurements.**
**(A, B)** NDO reconstruction fidelity as a function of the number of time steps for Hadamard QW with $\alpha \equiv \pi/4$ (green solid line), coherent disordered QW with $\alpha_t$ being time-dependent (red dashed line), and open QW with arbitrary mixing (blue dash-dotted line). The insets show the error in the purity of the reconstructed states $\rho_{\lambda\mu}$ for each time step. The shaded regions for the three types of QW are the standard errors of NDO reconstruction with 100 random samples for each step, and the lines are the averaging results. **(C, D)** show the corresponding results for the MaxLik method.

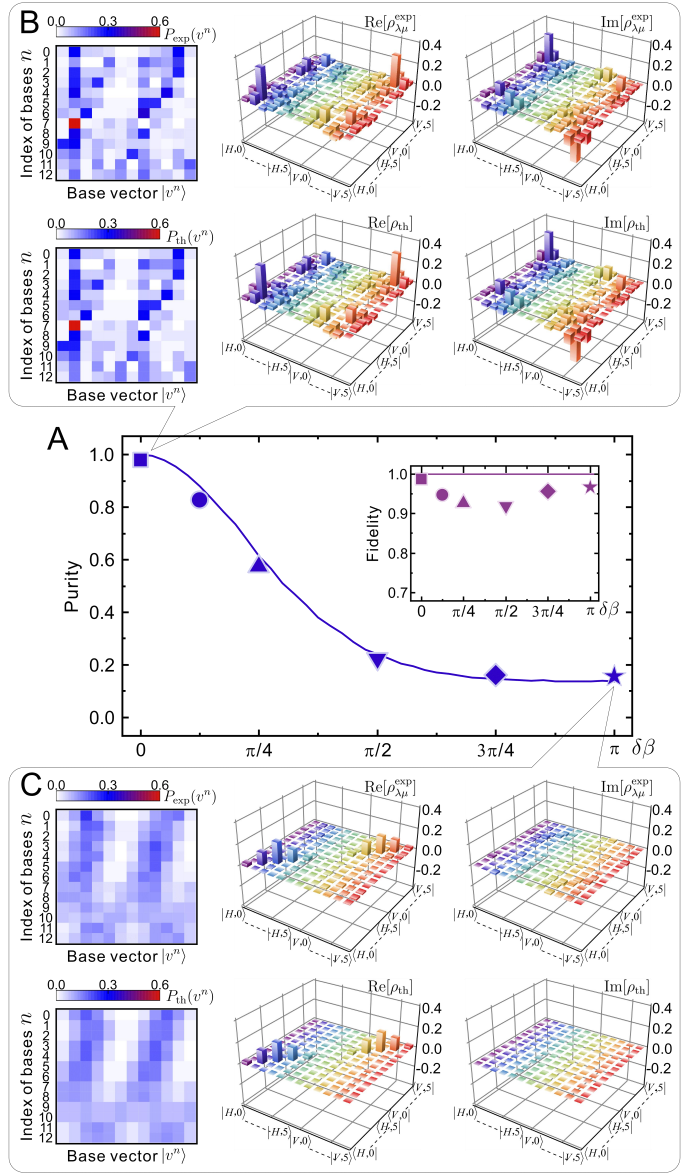

**Fig. 4. Experimental NDO tomography of open QW in a simulated environment.** **(A)** Purity of the NDO reconstructed states $\rho_{\lambda\mu}^{\mathrm{exp}}$ (blue symbols), trained on the experimental measurement data $P_{\mathrm{exp}}(v^n)$ for a five-step QW with six distinct mixings. The red symbols in the inset report the reconstruction fidelity. The blue and red solid lines give the theoretical results of target states $\rho_{\mathrm{th}}$. Error bars considering the statistical noise are smaller than the symbol size. **(B)** and **(C)** show the measured probability distributions $P_{\mathrm{exp}}(v^n)$ on 13 sets of bases, real and imaginary parts of $\rho_{\lambda\mu}^{\mathrm{exp}}$ (from left to right) for the Hadamard QW with $\delta\beta = 0$ and the fully decoherent walk with $\delta\beta = \pi$, respectively. The lower panels in **(B)** and **(C)** represent the corresponding theoretical expectations.

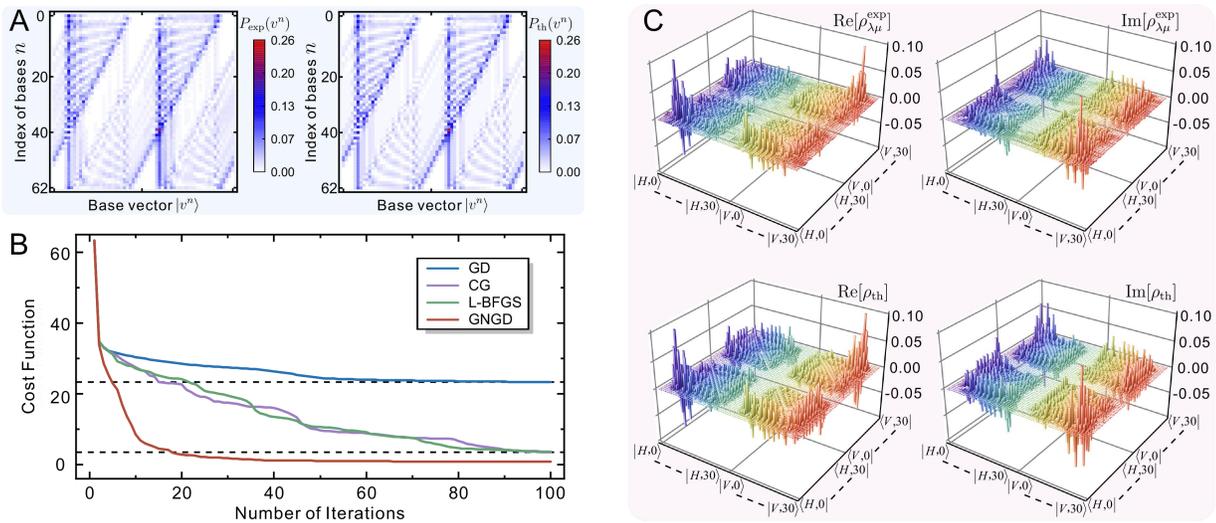

**Fig. 5. Experimental GNGD-enhanced NDO tomography of QW in a real environment.** **(A)** Measured probability distributions $P_{\text{exp}}(v^n)$ on 63 sets of bases for a 30-step Hadamard walk. The right panel is the theoretical expectation. **(B)** Cost function $\mathcal{D}_\theta$ versus the number of training iterations for NDO tomography using GD, CG, L-BFGS, and GNGD optimizer. **(C)** Real and imaginary parts of the NDO reconstructed state $\rho_{\lambda\mu}^{\text{exp}}$, and the theoretical expectations are shown in the lower panels.